\documentclass[nato,noid,numreferences]{crckbked}

\usepackage{graphicx}
\usepackage{epsf}
\begin{document}
% The \begin{document} command comes before the \begin{opening}
% command.

\begin{opening}
\title{CENTER VORTEX MODEL FOR NONPERTURBATIVE \protect\\
STRONG INTERACTION PHYSICS}

%\subtitle{Basic Instructions}
% Uncomment if you want to give a subtitle.

% You can split the title and subtitle by putting 
% two backslashes at the appropriate place. 

\author{Michael Engelhardt\thanks{Supported by DFG 
under grants Re 856/4-1 and Al 279/3-3.} }
\institute{Institut f\"ur Theoretische Physik, Universit\"at T\"ubingen\\
Auf der Morgenstelle 14, 72076 T\"ubingen, Germany}

\begin{abstract}
A model for the infrared sector of $SU(2)$ Yang-Mills theory, based on
magnetic vortex degrees of freedom represented by (closed) random
world-surfaces, is presented. The model quantitatively describes both the
confinement properties (including the finite-temperature transition to a
deconfined phase) and the topological susceptibility of the Yang-Mills
ensemble. A (quenched) study of the spectrum of the Dirac operator
furthermore yields a behavior for the chiral condensate which is
compatible with results obtained in lattice gauge theory.
\end{abstract}

\end{opening}

\section{Introduction}
\label{intro}
Strong interaction physics is characterized by diverse nonperturbative
phenomena. Color charge is confined, chiral symmetry is spontaneously
broken, and the axial $U(1)$ part of the flavor symmetry exhibits an
anomaly. Moreover, at finite temperatures, one expects to encounter a
deconfining phase transition. In principle, a theoretical tool exists
which permits the calculation of any observable associated with these
phenomena, namely lattice gauge theory.
Nevertheless, it is useful to concomitantly formulate effective models
which concentrate on the relevant infrared degrees of freedom, and thus
provide a clearer picture of the dominant physical mechanisms. This
facilitates the exploration of problem areas which are difficult to 
access using the full (lattice) gauge theory.
The vortex model \cite{selprep}-\cite{prepcs} presented in
the following aims to provide a comprehensive quantitative description
of all of the aforementioned nonperturbative aspects of the strong
interaction within a unified, consistent framework.
This turns out to be possible on the basis of a simple effective
model dynamics; namely, vortex world-surfaces can be regarded as
random surfaces on large length scales.

\section{Chromomagnetic center vortices and their dynamics}
\label{defsec}
Center vortices are closed lines of chromomagnetic flux in three-dimensional
space; thus, they are described by closed two-dimensional world-surfaces
in four-dimensional space-time. Their magnetic flux is quantized such that 
they contribute a phase corresponding to a nontrivial center element
of the gauge group to any Wilson loop they are linked to (or, equivalently,
whose minimal area they pierce). In the case of $SU(2)$ color discussed
here, the only such nontrivial center phase is $(-1)$. For higher gauge
groups, one must consider several possible fluxes carried by vortices.

\begin{figure}[h]
\epsfxsize=12cm
\epsffile{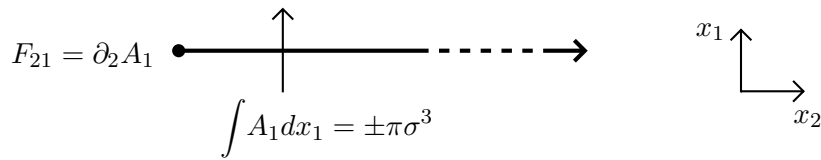}
\caption{Local construction of a gauge field describing a vortex.
Only the 1-2 plane is depicted; the vortex world-surface is a sheet
extending into the 3 and 4 directions. In the 1-2 plane, it therefore
shows up as a point. The vortex surface can be viewed as the
boundary of a three-dimensional volume in four-dimensional space-time;
in the two-dimensional slice of space-time depicted, this volume shows
up as a line emanating from the vortex. The volume can be used to
define the vortex gauge field: The latter shall have support precisely
on the volume (i.e.~be described by a $\delta $-function in the
coordinate locally perpendicular to the volume); as a Lorentz vector,
it shall point locally perpendicular to the volume, it shall point
in 3-direction in color space (as encoded in the third Pauli matrix
$\sigma^{3} $), and it shall have a magnitude such that the
line integral shown in the figure acquires the specified value.
Evaluating a Wilson loop enclosing the position of the vortex then yields
the value $-1$, as required. The field strength of the vortex is localized
on the vortex; its only nonvanishing tensor component is the one
associated with the two space-time directions locally perpendicular to
the vortex surface. Note that there is a free choice of sign of
the gauge field, corresponding to the two possible orientations of the
vortex flux. Note also that, globally, the gauge field support line in the
figure must end at another vortex elsewhere in the 1-2 plane; complications
can arise if the vortex surfaces (and the three-dimensional volumes spanning
them) are nonorientable. Such global issues are addressed in
section \ref{cssec}.}
\label{gfdef}
\end{figure}

As an illustrative example, consider a vortex surface located at a fixed
point on the 1-2 plane, i.e.~extending into the 3 and 4 directions. Such a
vortex surface can be associated with a gauge field which depends
only on the $x_1 $ and $x_2 $ coordinates as indicated in 
Fig.~\ref{gfdef}.

\vspace{-9.9cm}

\hspace{0.3cm} $F_{21} = \partial_{2} A_1 $

\[
\hspace{-2.1cm} \int \! A_1 dx_1 = \pm \pi \sigma^{3}
\]

\vspace{-2.2cm}

\hspace{9.4cm} $x_1 $

\vspace{0.7cm}

\hspace{10.7cm} $x_2 $

\vspace{8.9cm}

Note that the vortex in Fig.~\ref{gfdef} is infinitely thin, and the
associated field strength is concentrated on the thin vortex world-surface.
The real physical vortex fluxes conjectured to describe the infrared
aspects of the Yang-Mills ensemble within the vortex picture of course
should be thought of as possessing a finite
thickness\footnote{Phenomenologically, such a physical thickness
has e.g.~been argued to be crucial for an explanation of the Casimir
scaling behavior of adjoint representation
Wilson loops \cite{casscal}.}; however, for the purpose of evaluating 
deeply infrared observables, i.e.~looking from far away, the thin 
idealization is adequate. Nevertheless, the thickness of the vortices
will significantly influence the ansatz for the vortex dynamics presented
below, and the physical interpretation of that ansatz.

It should be remarked that the chromomagnetic vortex fluxes described above
correspond to the flux domains found in the so-called Copenhagen 
vacuum \cite{spag}. Thus, the idea that these degrees of freedom may be 
relevant in the infrared sector of Yang-Mills theory is not new. What is new 
about the model to be discussed here is that the vortex picture has been 
developed into a quantitative tool, allowing to evaluate diverse physical 
observables, and that it furthermore has been generalized to 
finite temperatures, including the deconfined phase.

To arrive at a tractable dynamics for the model vortex world-surfaces, the
latter will be composed of elementary squares (plaquettes) on a hypercubic
lattice. The spacing of this lattice will be a fixed physical quantity
related to the thickness of the vortex fluxes already mentioned further
above; the lattice prevents an arbitrarily close packing of the vortices. 
Thus, it is not envisaged to eventually take the lattice spacing to zero, 
and accordingly renormalize the coupling constants, such as to arrive at a 
continuum theory. Rather, the lattice spacing represents the fixed physical
cutoff one expects to be present in any infrared effective theory,
and thus also delineates the ultraviolet limit of validity of the model.
If one wants to refine the model such as to eliminate the artificial
hypercubic nature of the vortex surfaces, one has to replace the lattice
spacing by some other ultraviolet cutoff. For instance, if the surfaces
are represented as triangulations, a minimal area of the elementary
triangles could take on this role.

On the hypercubic lattice adopted here, the vortex surfaces will be
regarded as random surfaces. These surfaces will be generated using Monte
Carlo methods; the weight function specifying the ensemble depends on
the curvature of the surfaces as follows \cite{selprep}. Every instance of
a link on the lattice being common to two plaquettes which are part of a
vortex surface, but which do not lie in the same plane, is penalized by an
action increment $c$. Thus, the action can be represented pictorially as

\centerline{
\hspace{2.35cm}
\epsfxsize=1cm
\epsffile{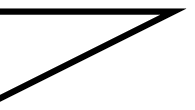}
}
\vspace{-1.58cm}
\begin{equation}
S_{curv} = c \cdot \mbox{ {\LARGE \# ( \ \ \ ) } }
\label{actio}
\end{equation}
Note that several such pairs of plaquettes can occur for any given
link. E.g., if six vortex plaquettes are attached to a link,
the action increment is $12c$.

\section{Confinement and Deconfinement}
Given the random surface dynamics defined in the previous section, it is
now straightforward to evaluate Wilson loops, using the fundamental
property of vortices that they modify any Wilson loop by a phase factor
$(-1)$ whenever they pierce its minimal area. Such measurements can
furthermore be carried out at several temperatures, by adjusting the
extension of (Euclidean) space-time in the time direction; this extension
is identified with the inverse temperature of the ensemble. At finite
temperatures, static quark potentials are given by Polyakov loop correlators;
their properties in the presence of vortex fluxes are completely 
analogous to the properties of Wilson loops. Qualitatively,
as long as the curvature coefficient $c$ is not too large, one finds a 
confined phase (non-zero string tension) at low temperatures, and a phase 
transition to a high-temperature deconfined phase. In order to make the
correspondence to full $SU(2)$ Yang-Mills theory quantitative, one can
adjust $c$ such as to reproduce the ratio of the deconfinement
temperature to the square root of the zero-temperature string tension,
$T_C /\sqrt{\sigma_{0} } =0.69$. This happens at the value $c=0.24$.
Furthermore, by setting $\sigma_{0} =(440 \, \mbox{MeV} )^2 $ to fix
the scale, one extracts from the measurement of $\sigma_{0} a^2 $ the
lattice spacing $a=0.39 \, \mbox{fm} $. As discussed in the previous
section, this is a fixed physical quantity, related to the thickness
of the vortices, which represents the ultraviolet limit of validity
of the effective vortex model.

\begin{figure}[ht]
\centerline{
\epsfxsize=7cm
\epsffile{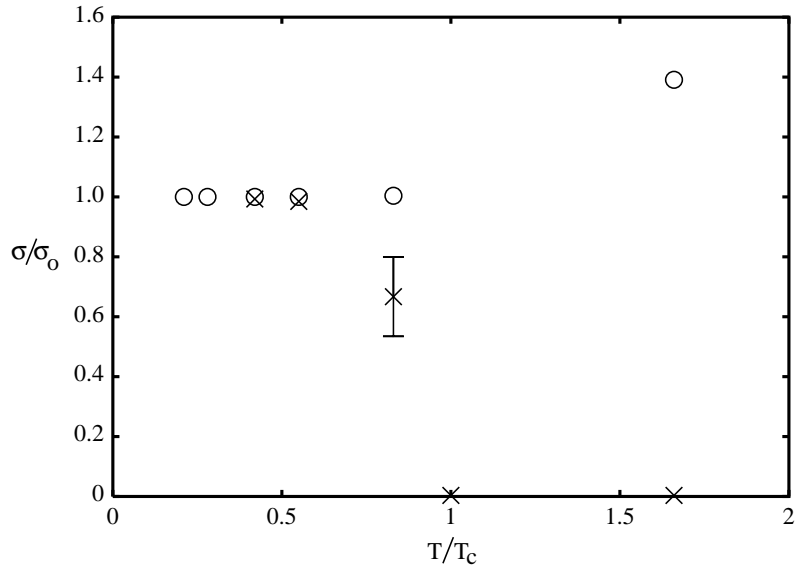}
\vspace{-2.1cm}
}
\caption{String tension between static color sources (crosses) and
spatial string tension (circles) as a function of temperature, obtained
with a curvature coefficient of $c=0.24$ on $16^3 \times N_t $ lattices.}
\label{stt}
\end{figure}

Fig.~\ref{stt} displays the results of string tension measurements on
$16^3 \times N_t $ lattices, as a function of temperature.
After having fixed $c$ as described above, the so-called spatial string
tension $\sigma_{s} $ can be predicted, cf.~Fig.~\ref{stt}. In the
high-temperature regime, it begins to rise with temperature; the
value obtained at $T=1.67 \, T_C $, namely 
$\sigma_{s} (T=1.67 \, T_C ) = 1.39 \sigma_{0} $, corresponds to
within 1\% with the value measured in full $SU(2)$ Yang-Mills 
theory \cite{karsch}.

\begin{figure}[h]
\centerline{
\epsfxsize=5cm
\epsffile{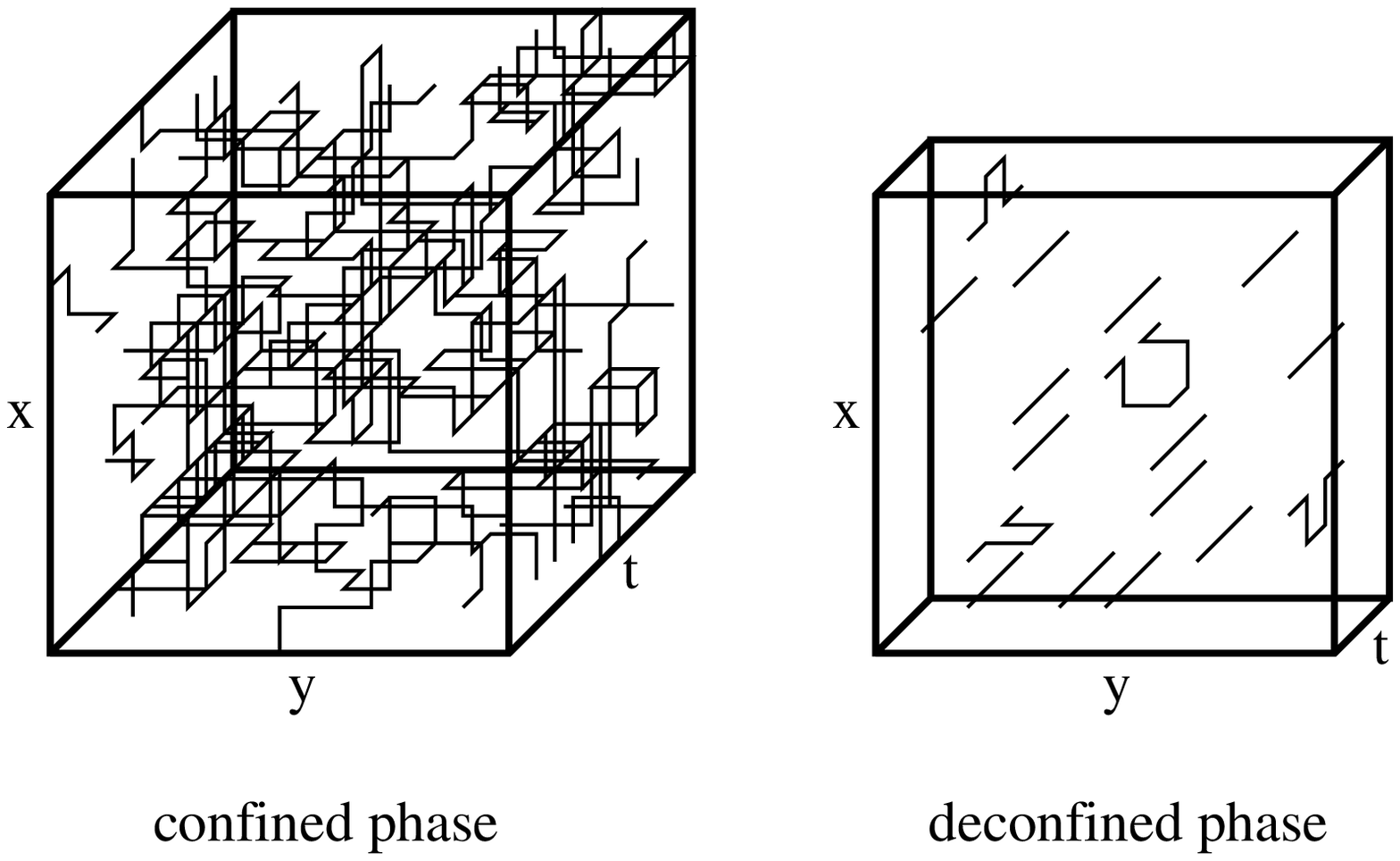}
\vspace{0.7cm}
}
\caption{Typical vortex configurations in the confined and the deconfined
phases.}
\label{pha}
\end{figure}

The confined and deconfined phases can alternatively be characterized by
the percolation properties \cite{selprep} of the vortex clusters in space 
slices of the lattice universe, i.e. slices defined by keeping one space 
coordinate fixed. In such a slice of space-time, vortices are represented 
by closed lines, cf. Fig. \ref{pha}. In the confined phase, these lines
percolate throughout (sliced) space-time, whereas in the deconfined phase,
they form small, isolated clusters, which, more specifically, wind around
the universe in the Euclidean time direction (and are closed by virtue of
the periodic boundary conditions). Also, simple heuristic arguments can be
given \cite{selprep} which explain why confinement should be associated 
with vortex percolation. The percolation characteristics of the surfaces 
in the vortex model closely mirror the ones found for vortex structures 
extracted from full lattice Yang-Mills configurations via an appropriate
gauge fixing and projection procedure \cite{deb97},\cite{giedt},
which were investigated in \cite{tlang}.

\section{Topology}
\label{topsec}
Besides confinement, the topological properties of the 
Yang-Mills ensemble constitute an important nonperturbative aspect of
strong interaction theory. These properties are encoded in the
topological charge
\begin{equation}
Q=\frac{1}{32\pi^{2} } \int d^4 x \, \epsilon_{\mu \nu \lambda \tau } \
\mbox{Tr} \ F_{\mu \nu } F_{\lambda \tau } \ .
\label{qdef}
\end{equation}
In view of (\ref{qdef}), nonvanishing topological density is generated at
a given space-time point if the field strength there has nonvanishing
tensor components such that the corresponding Lorentz indices span all
four space-time directions. For example, a nonvanishing $F_{12} $ in
conjunction with a nonvanishing $F_{34} $ fulfils this requirement.
Nonvanishing $F_{12} $ arises when a vortex surface segment (locally) runs
in 3-4 direction, as discussed in section \ref{defsec}; conversely, for
nonvanishing $F_{34} $, one needs a surface segment running
in 1-2 direction. Thus, a nontrivial contribution to the topological 
charge is e.g.~generated by a self-intersection point of the 
vortex surfaces, where a segment running in 3-4 direction intersects
a segment running in 1-2 direction. Quantitatively \cite{cont}, the 
contribution specifically of such a self-intersection point is
$\pm 1/2$. In general, all singular points of a surface configuration on
the hypercubic lattice contribute to the topological charge, where a
singular point is defined as a point (lattice site) at which the set of
tangent vectors to the surface
configuration spans all four space-time directions. Self-intersection
points are but the simplest example of such singular points; there
also exist writhings, at which the surface is twisted in such a
way as to generate a singular point in the above sense. These writhings
actually turn out to be statistically far more important than the
self-intersection points in the random surface ensemble discussed 
here \cite{preptop}.

Contrary to the Wilson loop, the topological charge $Q$ is sensitive
to the orientation of the vortex surfaces via the signs of the field
strengths entering (\ref{qdef}). The random surfaces of the vortex ensemble
should not be thought of as being globally oriented (in fact, most
configurations are not even globally orientable); they in general are
composed of patches of alternating orientation. Given a vortex surface
from the ensemble defined by (\ref{actio}), it is straightforward to
furthermore randomly assign
orientations to the plaquettes making up the surface. By biasing this
procedure with respect to the relative orientation of neighboring
plaquettes, different mean sizes of the oriented patches making up
the surface can be generated; equivalently, the density of patch
boundary lines can be adjusted. This density strictly speaking
constitutes an additional parameter of the model, which cannot be fixed
using the confinement properties due to the fact that the Wilson loop is
insensitive to the vortex orientation. A priori, one might expect e.g.~the
topological susceptibility $\chi = \langle Q^2 \rangle /V$, where $V$
denotes the space-time volume under consideration, to depend on this
parameter. This would mean that $\chi $ can possibly be fitted, but not
predicted. In actual fact, it turns out that this quantity is
independent of the aforementioned density within the error bars. The 
reasons for this can be understood in detail in terms of the geometrical
properties of the random surfaces \cite{preptop}. The measurement
of the topological susceptibility $\chi $ discussed below thus does
represent a genuine quantitative prediction of the vortex model.

\begin{figure}[t]
\centerline{
\epsfxsize=11cm
\epsffile{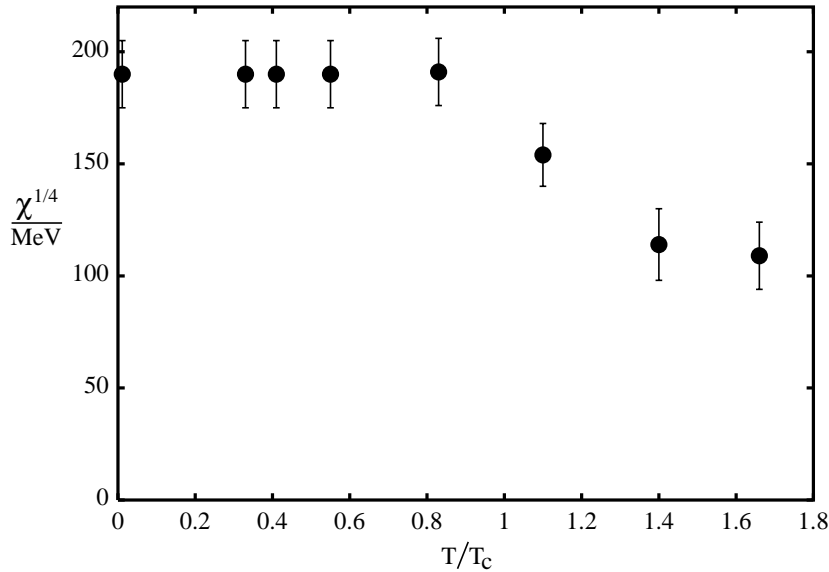}
}
\caption{Fourth root of the topological susceptibility as a function of
temperature, obtained on $12^3 \times N_t $ lattices.}
\label{chifig}
\end{figure}

In practice, before the topological charge can be extracted, in the manner
indicated above, from the geometrical properties of the model hypercubic
lattice surfaces, ambiguities must be resolved \cite{preptop}
which would not appear if one were dealing with arbitrary surfaces in
continuous space-time, and which are reminiscent of the ones occuring in
standard lattice gauge theory. Only after construction of an appropriate
``inverse-blocking'' algorithm, the topological charge
$Q$ of the vortex surface configurations can be evaluated, allowing to
determine the topological susceptibility of the vortex ensemble.
The result is exhibited in Fig. \ref{chifig} as a function of temperature.

Quantitatively, this result is compatible with measurements in full
Yang-Mills theory \cite{digia}. The vortex model thus provides, within
one common framework, a simultaneous, consistent description of both the
confinement properties as well as the topological properties of the
$SU(2)$ Yang-Mills ensemble. It should be remarked that, also as far as the 
vortex structures extracted from lattice Yang-Mills configurations via
center gauge fixing and center projection \cite{deb97},\cite{giedt} are
concerned, evidence exists that these structures encode
the topological characteristics of the gauge fields
\cite{forc1},\cite{protop}.

\section{Spontaneous chiral symmetry breaking}
\label{cssec}
A comprehensive description of the nonperturbative phenomena which
determine strong interaction physics must furthermore include the
coupling of the vortices to quark degrees of freedom and the associated
spontaneous breaking of chiral symmetry. The latter can be quantified via
the chiral condensate $\langle \bar{\psi } \psi \rangle $, which is
related to the spectral density $\rho (\lambda )$ of the Dirac operator
$D \! \! \! \! / \, $ in a vortex background via the Casher-Banks formula
\cite{cashb}
\begin{equation}
\langle \bar{\psi } \psi \rangle = -\lim_{V\rightarrow \infty }
2m\int_{0}^{\infty } d\lambda \, \frac{\rho (\lambda )}{m^2 +\lambda^{2} }
\ ,
\label{cabaf}
\end{equation}
where $m$ denotes the quark mass and $V$ the space-time volume. In the
chiral limit $m\rightarrow 0$, the chiral condensate thus behaves as
$\langle \bar{\psi } \psi \rangle \rightarrow -\pi \rho (0)$.

\begin{figure}[h]
\centerline{
\hspace{-0.7cm}
\epsfxsize=6.5cm
\epsffile{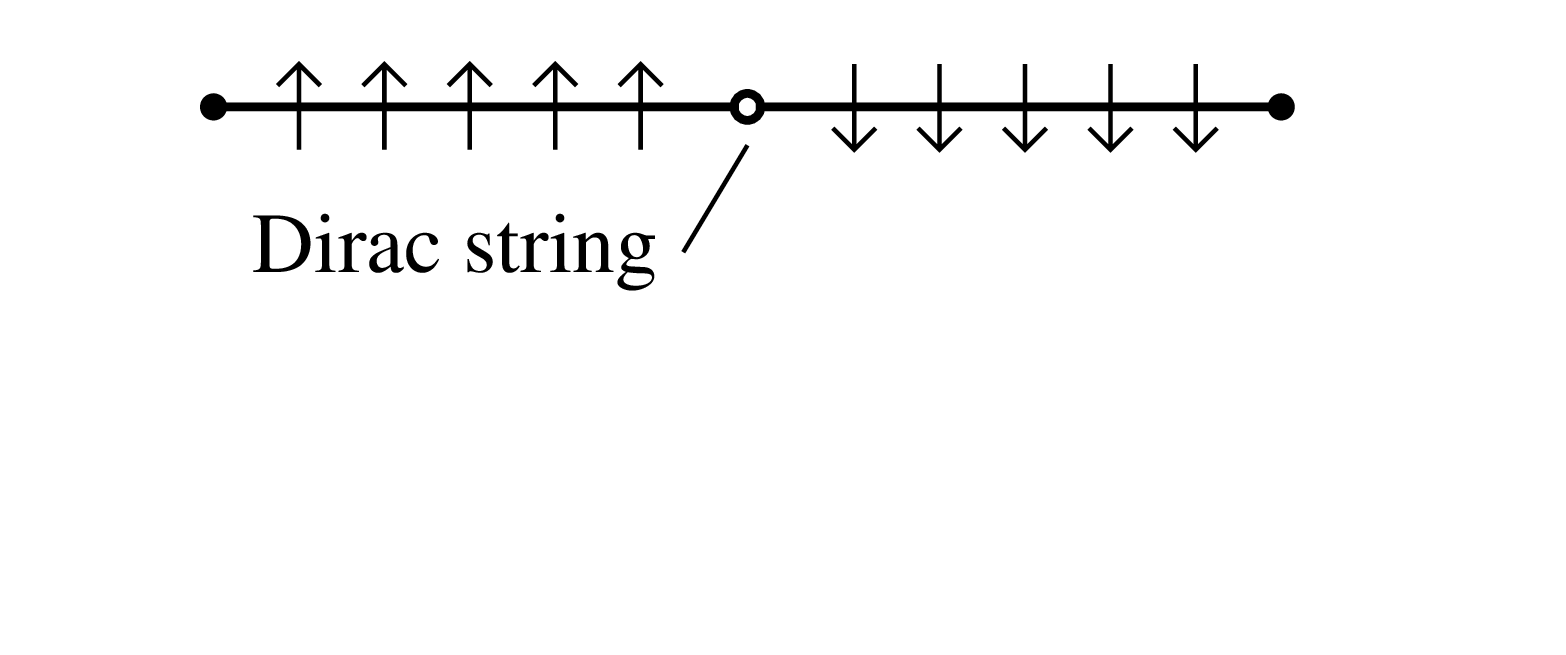}
\hspace{-1.1cm}
\epsfxsize=6.5cm
\epsffile{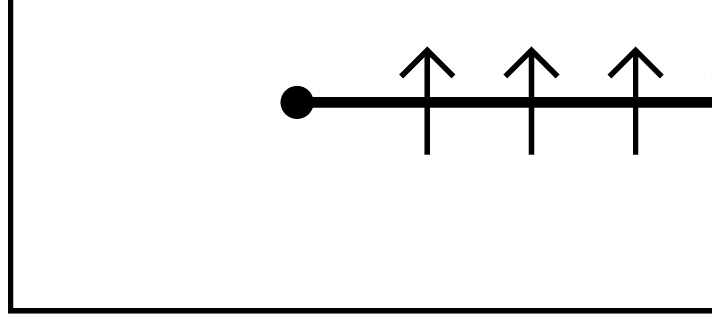}
}
\caption{Nonoriented gauge field support volume in a two-dimensional slice
of space-time. In such a slice, vortices show up as the solid points,
volumes emanating from them as the solid lines. The directions of the
vortex fluxes at the boundaries of the volumes, fixed a priori, here
happen to be such that the volume segments emanating from the different
vortices are forced to have mutually incompatible orientation. This is
indicated by the arrows, which symbolize (directed) line integrals
$\int A_{\mu } dx_{\mu } $; {\em all of the line integrals indicated in
the figures yield the same value}, for definiteness $+\pi \sigma^{3} $.
Left: Attempt at a global definition of the gauge field; the gauge field
support volume then necessarily contains the indicated Dirac string,
carrying flux double that of a vortex. Right: Wu-Yang construction, with
space-time subdivided into two patches, on each of which the volume is
oriented. Gauge fields on the two patches are related by the transition
function $U$. In the simple example depicted here, $U$ can be chosen
constant in the whole overlap region shared by the two patches. In generic
configurations, the presence of other vortices nearby forces the space-time
region of nontrivial $U$ to be more localized, namely onto the immediate
vicinity of the gauge field support volume. For further details on this
point and the complications it entails, cf.~\cite{prepcs}.}
\label{fig1}
\end{figure}

In order to construct the Dirac operator, it is necessary to cast the
vortex fluxes explicitly in terms of gauge fields. Locally, this can
be achieved as displayed in Fig.~\ref{gfdef}, but globally, a difficulty
arises: Generic vortex surfaces are not orientable. Viewing
vortex surfaces as boundaries of three-volumes (which represent the
support of the associated gauge field, cf.~Fig.~\ref{gfdef}), also these
three-volumes are in general not oriented.
As a consequence, the gauge field either contains Dirac strings or
must be defined on different space-time patches in the spirit of the
Wu-Yang construction, cf.~Fig.~\ref{fig1}. In practice, the description
via Dirac strings is not suitable, for the following reason:
If one solved the Dirac equation in the presence of Dirac strings exactly,
then the quark wave functions would exhibit singularities along the Dirac
strings which would cancel any physical effect of these strings.
The Dirac strings would be unobservable, as they should be.
However, in a truncated calculation, employing only infrared quark
modes, cf.~below, the cancellation would not be
perfect; instead, the Dirac strings effectively would act as additional
physical magnetic fluxes (of magnitude double that of vortex fluxes)
with which the quarks can interact. Thus, more magnetic disorder would
in effect be present than the model aims to describe.
Consequently, it is more appropriate to use the Wu-Yang construction
for the gauge field. On the individual space-time patches, which must be
chosen sufficiently small, the vortices are orientable and no Dirac
strings arise. The patches then are related by transition functions,
which are chosen non-Abelian, cf.~Fig.~\ref{fig1}.

Having defined an explicit gauge field representation of the vortex
configurations, one can evaluate (analytically) all matrix elements
of the Dirac operator in a truncated, infrared basis of quark wave
functions. This matrix representation of the Dirac operator then
determines the propagation of the infrared quark modes spanned by
the basis. The basis used in the present work is of the finite element
type \cite{prepcs}; for each $2^4 $ cube in the hypercubic lattice on
which the vortex surfaces are defined, there is a quark basis function
which is localized on that cube (and which is piecewise linear in each of
the space-time coordinates). These $2^4 $ cubes serve a dual purpose;
they at the same time represent the space-time patches which are used
in practice to define the vortex gauge field as discussed further
above.

Before turning to the numerical results, note that some options still
remain within the above construction scheme for the gauge field: Since the
gauge field support three-volume only must satisfy the property that its
boundary reproduces the locations of the vortex fluxes, one can freely
choose its interior. It can be constructed
in a random fashion, which will yield a rather rough three-volume, or
in a smoothed fashion. Likewise, when defining the gauge fields on the
individual space-time patches, one can choose a procedure which is not
biased with respect to the relative color orientations of adjacent
patches; such a random relative color orientation will yield a high
density of nontrivial transition functions. Alternatively, one can
consider supplementing this by (gauge) transformations $A\rightarrow -A$ on
individual patches such as to align their color orientations with those
of neighboring patches, leading to maximally smooth transition functions.

\begin{figure}[t]
\centerline{
\epsfxsize=11cm
\epsffile{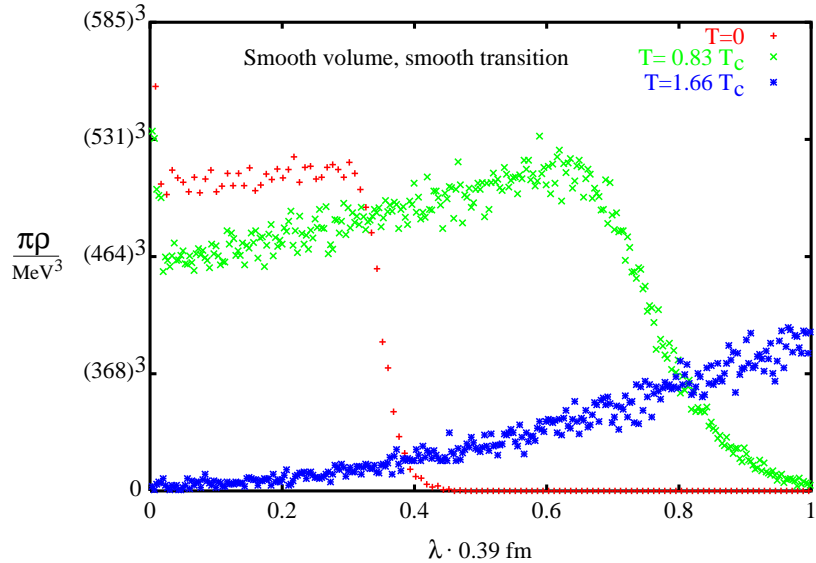}
}
\caption{Dirac spectral density at different temperatures for the
maximally smooth construction of the vortex gauge field, obtained using
a spatial extension of the lattice of $4a=1.56\, $fm.}
\label{csbfig}
\end{figure}

Having constructed a finite matrix representation of the Dirac operator in
the above fashion, one can numerically evaluate the (vortex ensemble average
of the) Dirac spectral density $\rho (\lambda )$. The results for the
maximally smooth gauge field construction (in the sense of the previous
paragraph) are depicted in Fig.~\ref{csbfig}. The qualitative properties of
the spectrum of the Dirac operator are as follows. At very small eigenvalues
$\lambda $, there is an anomalous enhancement of the spectral density which
is presumably due to the quenched approximation used here; a similar effect
has e.g.~been argued to occur in quenched chiral perturbation theory
\cite{chipd} and may contribute to the divergent chiral condensate in
the chiral limit observed in lattice calculations employing domain wall
fermions \cite{chen}. Apart from this divergence, the bulk of the spectrum,
which for sufficiently large quark masses dominates the value of the
chiral condensate, cf.~eq.~(\ref{cabaf}), extrapolates to a finite value
at $\lambda =0$ in the confining phase; this signals the spontaneous
breaking of chiral symmetry. Quantitatively, the zero-temperature chiral
condensate extracted by extrapolating the bulk of the spectrum to zero
eigenvalue $\lambda $, cf.~Fig.~\ref{csbfig}, is of the
same order of magnitude as the one found in lattice Yang-Mills
theory \cite{hatep}, which amounts to $-(450\, \mbox{MeV})^{3} $ (the scale
having been fixed in the same way as in the present model). This result is
reasonably robust if one uses rougher gauge field constructions (in the
sense discussed above) \cite{prepcs}; the result for the zero-temperature
chiral condensate then varies between $-(465\, \mbox{MeV})^{3} $ and
$-(560\, \mbox{MeV})^{3} $. It should be noted that the actual value
of the condensate by itself carries no direct physical meaning, since it
is not a renormalization group invariant; only its product with the quark
mass (which is as yet undefined within the model) is. However, the order of
magnitude agreement with lattice gauge theory indicates that the 
vortex model yields a natural description of spontaneous chiral
symmetry breaking, without requiring e.g.~unnaturally large or small
quark masses. More detailed insight into the quark physics induced
by the vortex ensemble will require the evaluation of further hadronic
observables.

Turning to the deconfined phase, the differences between the various model
options for the gauge field mentioned further above by contrast already
become apparent at the qualitative level. The smoothest option, which
is displayed in Fig.~\ref{csbfig}, is the only one which reproduces the
behavior found in lattice gauge theory, namely a rapid drop in the chiral
condensate as the temperature is raised above the deconfining phase
transition. This appears to represent the model of choice. All other,
rougher, gauge field models \cite{prepcs} retain a substantial chiral
condensate in the deconfined phase. It thus seems crucial to remove,
as much as possible, ultraviolet artefacts in the vortex gauge fields in
order to correctly describe that phase.

\section{Outlook}
\label{outsec}
With the Dirac operator at hand, one can also envisage carrying out
dynamical quark calculations within the vortex model, by reweighting
the vortex ensemble with the Dirac operator determinant. This will
substantially penalize Dirac eigenvalues of very small magnitude and
thus presumably reduce the chiral condensate to the phenomenologically
expected values around $-(230\, \mbox{MeV})^3 $. Moreover, to make contact
with phenomenology, the vortex model must still be extended to $SU(3)$
color; the corresponding random surface ensemble will in some respects be
qualitatively different from the one discussed here, since there are two
nontrivial center elements in the $SU(3)$ group, namely the phases
$e^{\pm i 2\pi /3} $. One therefore must allow for
two distinct vortex fluxes, which can branch and fuse
into one another. This difference in the topological character of the
configurations is e.g.~expected to lead to a change in the order of the
deconfinement phase transition from second order for $SU(2)$ to first
order for $SU(3)$, as observed in full Yang-Mills lattice experiments
\cite{kopo}. Ultimately, it is hoped that this
model will become a useful tool for phenomenological considerations.

\end{document}